\documentclass[twocolumn]{aastex631}

\usepackage{times}

\usepackage{graphicx}	
\usepackage{amsmath}	
\usepackage{amssymb}	

\newcommand{\bz}{\langle B_z \rangle}

\newcommand{\halp}{H$\alpha$}

\newcommand{\hd}{HD\,54879\,}

\begin{document}
\title{
Characterizing the dynamical magnetosphere of the extremely slowly rotating
magnetic O9.7\,V star HD\,54879 using rotational modulation of the H$\alpha$
 profile
}



\author{M. K\"uker}
\affiliation{Leibniz-Institut f\"ur Astrophysik Potsdam (AIP), An der Sternwarte 16, 14482 Potsdam, Germany}
\author{S. P.~J\"arvinen}
\affiliation{Leibniz-Institut f\"ur Astrophysik Potsdam (AIP), An der Sternwarte 16, 14482 Potsdam, Germany}
\author{S. Hubrig}
\affiliation{Leibniz-Institut f\"ur Astrophysik Potsdam (AIP), An der Sternwarte 16, 14482 Potsdam, Germany}
\author{I. Ilyin}
\affiliation{Leibniz-Institut f\"ur Astrophysik Potsdam (AIP), An der Sternwarte 16, 14482 Potsdam, Germany}
\author{M. Sch\"oller}
\affiliation{European Southern Observatory, Karl-Schwarzschild-Str. 2, 85748 Garching, Germany}






\begin{abstract}
The magnetic field in the O9.7\,V star HD54879 has been monitored for almost a 
decade. Spectropolarimetric observations  reveal a rather strong mean 
longitudinal magnetic field that varies with a period of about 7.41\,yr. 
Observations  in the H$\alpha$ line show a  variation with the same period, 
while the H$\beta$ line shows only little variation. Assuming the periodic  
variation to be caused by a slow rotation and a dipolar magnetic field, we 
find a magnetic field strength of $\ge$2\,kG at  the magnetic poles. With 
the relatively low mass loss rate of $10^{-9} M_\odot$\,yr$^{-1}$, this 
star is a case of extremely strong magnetic confinement. Both theoretical 
arguments and numerical simulations indicate the presence of an extended disk 
of increased gas density in the equatorial plane of the magnetic field, where 
gas from the line-driven stellar wind is trapped. This disk is likely to be 
the origin of the observed H$\alpha$ emission, which peaks together with the 
strongest line-of-sight magnetic field. The profile of the H$\alpha$ line is 
resolved in several components and shows a remarkable variability with the 
rotation period.
\end{abstract}

\keywords{stars: individual: HD\,54879 -- stars: early-type -- stars: atmospheres -- stars: variables: general -- stars: magnetic fields }


\section{Introduction}
\label{sec:intro}
The role of massive stars in the evolution of our Universe is widely 
recognized: massive stars drive energetic physical processes that affect 
the structure of entire galaxies and chemically enrich the ISM. Magnetic fields
are considered to be key components of massive stars with a far-reaching 
impact on their evolution and ultimate fate: a magnetic mechanism for the 
collimated explosion of massive stars, relevant for long-duration gamma-ray 
bursts, X-ray flashes, and asymmetric core collapse supernovae was proposed by 
\cite{Uzdensky2006}.
Magnetic O-type stars with masses larger than 30\,$M_{\odot}$ and their WR 
descendants are potentially the progenitors of magnetars 
\citep{Gaensler2005}.
Magnetars may be the origin of the powering mechanism in both superluminous 
supernovae and long-duration gamma-ray bursts 
\citep[e.g.][]{Yu2017}. 
Merging of binary compact remnants produces astrophysical transients
detectable by gravitational wave observations 
\citep[e.g.][]{abbott21}.

However, the origin of the magnetism in massive stars and the role the 
magnetic field plays in their evolution remain unknown. The currently most 
popular scenario involves a merging event, or mass transfer, or common 
envelope evolution 
\citep[e.g.][]{Tout2008, Ferrario2009, Schneider2016}. 
Mass transfer or stellar merging rejuvenates the mass gaining star, while 
the induced differential rotation is thought to be the key ingredient to 
generate a magnetic field 
\citep[e.g.][]{Wickramasinghe2014}.

Previous spectropolarimetric surveys of massive stars indicated that only 
one short-period binary system, Plaskett's star, contains a hot, massive, 
magnetic companion
\citep{Grunhut2013}.
On the other hand, a recent study of a representative sample of binary and 
multiple systems using the European Southern Observatory (ESO) High 
Accuracy Radial velocity Planet Searcher polarimeter
\citep[HARPS\-pol;][]{Snik2008}
and the Canada-France-Hawaii Telescope (CFHT) Echelle SpectroPolarimetric 
Device for the Observation of Stars 
\citep[ESPaDOnS;][]{Donati2006},
archival spectropolarimetric observations revealed a sizeable sample of 
binary and multiple systems with magnetic and potentially magnetic 
components 
\citep{Hubrig2023}.

\cite{Castro2015}
were the first to report a longitudinal magnetic field of $-600$\,G for
the presumably single O9.7\,V star \hd{}. Such a field yields to a lower 
limit of the dipole strength of about 2\,kG.
\cite{Hubrig2020} 
reported a possible increase in radial velocity by about 300\,m\,s$^{-1}$
between 2018 January and 2020 December. However, archival VLTI observations 
did not find any companion to \hd{} up to 3.5 mag fainter than the primary star 
\citep{Jarvinen2022}. 

The spectropolarimetric observations of \hd{} acquired over the last decade 
show a very slow magnetic field variability related to the extremely slow 
rotation of \hd{}, which is also indicated in a dynamical spectrum, 
displaying variability of the \halp{} line 
\citep[e.g.,][]{Jarvinen2022, Hubrig2020}. 
Assuming that the magnetic field of \hd{} has a pure dipolar configuration, 
\cite{Jarvinen2022} 
fitted a cosine curve to the observed distribution of data points obtained 
from the high-resolution spectropolarimetric observations and determined a
stellar rotation period of 7.2\,yr. The authors also reported on the 
anomalous element distribution in the atmosphere of this star: remarkable 
differences in the measurements using line masks for the elements O, Si, 
and He were detected on a few epochs close to the best visibility of the 
negative magnetic field pole. The field strengths measured from He lines 
were systematically the lowest whereas those from O lines were always higher 
than those measured from Si lines. Further, the numerical magnetospheric 
model presented in this work suggests the existence of an enhanced gas 
density that fills the volume inside the field lines close to the star.

In the following we present recent spectroscopic and polarimetric 
observations of  \halp{}, which together with previous observations now 
cover a time span of { 9.9}\,yr and thus a full rotation period. Based on 
numerical simulations of the wind originating in \hd{} using the Nirvana MHD 
code we present the distribution of mass density and gas velocity 
amplitude in its magnetosphere. As the magnetosphere is the source of the 
\halp{} emission observed in massive stars, we discuss the importance of 
the study of the rotational variability of this emission line to 
characterize the dynamical magnetosphere around this extremely slowly 
rotating magnetic star.


\section{Observations and mean longitudinal magnetic field measurements}
\label{sec:obs}

The observations of \hd{} presented in this work were obtained with multiple 
instruments. The FOcal Reducer low dispersion Spectrograph
\citep[FORS\,2;][]{Appenzeller1998},
the Ultraviolet and Visual Echelle Spectrograph 
\citep[UVES;][]{uves},
and HARPS\-pol
\citep{Snik2008}
are instruments offered by the ESO, the first two installed at the Very 
Large Telescope (VLT) on Cerro Paranal, Chile, and the latter one at ESO's 
3.6\,m telescope on La Silla, Chile. In addition, we have used in our 
analysis observations from the high-resolution Potsdam Echelle 
Polarimetric and Spectroscopic Instrument 
\citep[PEPSI;][]{PEPSI} installed
at the 2$\times$8.4\,m Large Binocular Telescope (LBT) in Arizona and 
archival spectropolarimetric observations recorded with 
the ESPaDOnS installed at the CFHT.

The majority of the observations and the data reduction was described in 
our previous work on \hd{}
\citep{Hubrig2020,Jarvinen2022}.
In the following sections only the previously unpublished observations are
discussed. 

\subsection{FORS\,2}

Two new spectropolarimetric observations of \hd{} with FORS\,2 were obtained 
on 2022 December~31 and 2023 October~8 in the framework of the ESO 
programmes 0110.D-0103(A) and 0112.D-2090(A), respectively. The observational 
setups were in both cases the same as presented previously by 
\cite{Hubrig2020}.
We always use a non-standard readout mode with low gain since it provides 
a broader dynamic range and leads to a higher signal-to-noise ratio 
($S/N$) in the individual spectra. He-Ne-Ar arc lamp exposures are 
obtained for wavelength calibration. The data extraction is described by
\citet{Cikota}.

\subsection{PEPSI}
\label{sec:PEPSIobs}

Circular polarized PEPSI data of \hd{} were obtained on 2022 January~10. For 
these observations, a different wavelength region was used in comparison 
to the observations described in 
\citet{Jarvinen2022}.
Using crossdispersers III and V, the obtained spectra cover the two wavelength 
regions 4765--5400\,\AA{} and 6271--7419\,\AA{}. The crossdisperser V was 
chosen to record the H$\alpha$ line. The observations and the data reduction 
followed the descriptions given by
\citet{Jarvinen2022}.
{
Additional observations of integral light spectra of HD\,54879 with PEPSI were obtained on 2023 December~3.
For these ovservations we used the crossdispersers
II, IV, and V covering the wavelength regions 4238--4768\,\AA{}, 5389--6283\,\AA{},
and 6271--7419\,\AA{}, respectively.
}
\subsection{ESPaDOnS}

Publicly available and already reduced ESPaDOnS polarimetric spectrum 
obtained on 2021 February~24 was downloaded from the CFHT archive.

\subsection{UVES}

In total three new spectra have been recorded with UVES mounted on UT2 of 
the VLT. The first two spectra were obtained in the framework of the 
programme 0108.D-0233(B) carried out in 2021 November~29 and 2022 March~1, 
and one more spectrum was obtained within the programme 0110.D-0103(B) on 
2023 March~31. Similar to the previously analysed UVES spectra in 
\citet{Jarvinen2022},
the obtained spectra have a resolving power $R\approx80,000$ in the blue arm  
(3756--4982\,\AA) and $R\approx110,000$ in the red arm (5690--9459\,\AA). 
The data was reduced with the ESO Phase~3 UVES 
pipeline\footnote{http://www.eso.org/rm/api/v1/public/releaseDescriptions/163}.

\subsection{HARPS\-pol}

One new spectropolarimetric observation of \hd{} was obtained on 2024 
January~1. Similar to previous HARPS\-pol data, the spectra have a resolving power 
of $R\approx115,000$ and cover the wavelength range 3780--6912\,\AA{} with 
a gap at 5259--5337\,\AA{}.
%
\begin{table}
\caption{Logbook of the 
so far unpublished high-resolution
observations analysed in this paper.
The first column gives the instrument used to obtain the spectrum, the
second column the time of the observation as modified Julian date 
(MJD), and the third column the $S/N$ measured near the H$\alpha$ line.
The $S/N$ value marked with an asterisk was determined around 5860\,\AA{}
because the H$\alpha$ line was not included in those observations.
The final column presents the results of the measurements of the longitudinal
magnetic field with the spectropolarimetric instruments.
}
\label{tab:hr_obs}
\begin{tabular}{cccr@{$\pm$}r}
\hline 
\hline 
\multicolumn{1}{c}{Instrument} &
\multicolumn{1}{c}{MJD} &
\multicolumn{1}{c}{$S/N$} &
\multicolumn{2}{c}{$\left< B_{\rm z} \right>$} \\
\multicolumn{1}{c}{} &
\multicolumn{1}{c}{} &
\multicolumn{1}{c}{} &
\multicolumn{2}{c}{[G]} \\
  \hline 
  ESPaDOnS & 59269.38 & 313 & $-$494 & 25 \\
  UVES     & 59548.34 & 264 & \multicolumn{2}{r}{--} \\
  UVES     & 59640.13 & 247 & \multicolumn{2}{r}{--} \\
  UVES     & 60035.09 & 190 & \multicolumn{2}{r}{--} \\
  PEPSI    & 59589.26 & 586 & $-$404 & 50 \\
  PEPSI    & 60281.51 & 411 & \multicolumn{2}{r}{--} \\
  HARPS    & 60311.08 & 129 & $-$361 &  7\\
\hline 
\end{tabular}
\end{table}
\begin{table}
\caption{
Logbook of the new low resolution FORS\,2 spectropolarimetric 
observations of \hd{}.
The first column gives the modified Julian dates (MJD) at the middle of 
the exposure and the second column the $S/N$ of the spectra, measured at 4800\,\AA{}. 
Longitudinal magnetic field measurements, those for the entire spectrum 
and those using only the hydrogen lines, are presented in Columns~3 and 4. 
The last column has the measurements using all lines in the null spectra, 
which are obtained from pairwise differences from all available Stokes~$V$ 
profiles so that the real polarization signal should cancel out.
}
\label{tab:hr_obs_fors}
\begin{tabular}{crr@{$\pm$}rr@{$\pm$}rr@{$\pm$}r}
\hline 
\hline 
\noalign{\smallskip}
\multicolumn{1}{c}{MJD} &
\multicolumn{1}{c}{$S/N$} &
\multicolumn{2}{c}{$\left< B_{\rm z} \right>_{\rm all}$} &
\multicolumn{2}{c}{$\left< B_{\rm z} \right>_{\rm hyd}$} &
\multicolumn{2}{c}{$\left< B_{\rm z} \right>_N$} \\
\multicolumn{1}{c}{} &
\multicolumn{1}{c}{} &
\multicolumn{2}{c}{[G]} &
\multicolumn{2}{c}{[G]} &
\multicolumn{2}{c}{[G]} \\
\noalign{\smallskip}
  \hline 
\noalign{\smallskip}
59945.35 & 1314 & $-$513 & 125 & $-$889 & 209 & 87 & 119 \\
60226.34 & 2020 & $-$573 & 69  & $-$895 & 142 & $-$32 & 66 \\
\hline 
\end{tabular}
\end{table}


\subsection{Longitudinal magnetic field measurements}

All {  new} high-resolution observations obtained with and without 
a Zeeman analyzer are listed in Table~\ref{tab:hr_obs}. The 
{  new measurements} obtained with FORS\,2 {  are} presented in 
Table~\ref{tab:hr_obs_fors}.

The low-resolution polarimetric spectra obtained with FORS\,2 and the 
high-resolution spectropolarimetric observations with ESPaDOnS, PEPSI and HARPS 
were used to determine the mean longitudinal magnetic field $\bz{}$.
The analysis of the ESPaDOnS, PEPSI, and HARPS\-pol observations, similarly 
to the work of
\citet{Hubrig2020}, 
is based on the least-squares deconvolution (LSD) technique
\citep[for details, see][]{Donati1997}.
For the ESPaDOnS observation obtained on 2021 February~24 we measure
$\bz=-494\pm25$\,G whereas for the PEPSI observation obtained on 2022 
January~10 we measure $\bz=-404\pm50$\,G. The HARPS\-pol observation
obtained on 2024 January~1 gives $\bz=-361\pm7$\,G. Using the false alarm 
probabilities (FAPs), following the definitions by 
\citet{Donati1992},
these measurements are definite detections with FAPs $< 10^{-6}$. 

The mean longitudinal magnetic field $\bz{}$ using the low-resolution FORS\,2 
observations obtained on 2022 December~31 and 2023 October 8 were measured 
following the description given by e.g.\
\cite{Hubrig2004a,Hubrig2004b}
and references therein. The error estimations were based on a Monte Carlo 
bootstrapping test 
\citep[e.g.][]{Steffen2014}. 
Similar to our previous studies
\citep[][and references therein]{Jarvinen2022},
the FORS\,2 field strengths obtained using exclusively hydrogen lines, 
$\bz=-889\pm209$\,G and $\bz=-895\pm142$\,G, are in absolute value higher 
than those obtained using the entire spectrum, $\bz=-513\pm125$\,G and 
$\bz=-573\pm69$\,G. 
%
\begin{figure}
\includegraphics[width=\columnwidth]{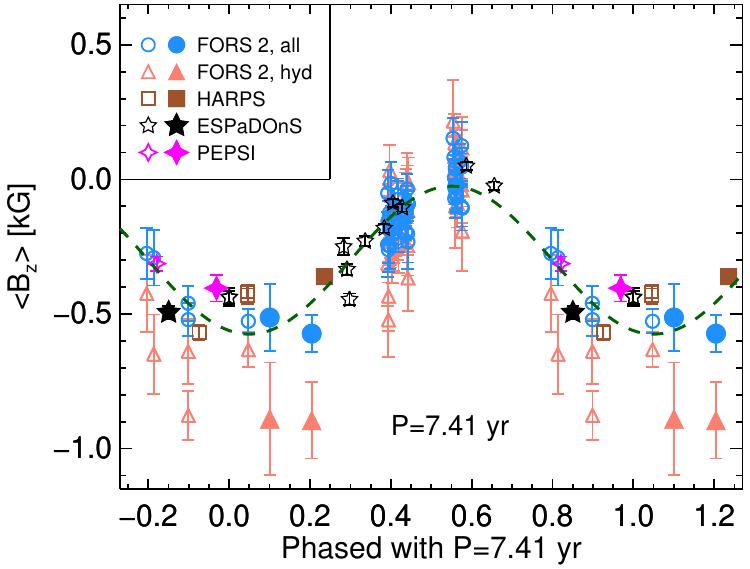}
\caption{ Distribution of the mean longitudinal magnetic field $\bz{}$ values
of \hd{} as a function of phase. The smaller {  open} symbols represent 
data that were published in
\cite{Jarvinen2022}.
Larger {  filled} symbols are used to present the new, not previously 
published, measurements. {  Circles indicate FORS\,2 measurements 
obtained using the entire spectrum and triangles} measurements where only 
the hydrogen lines were used. {  Squares, stars, and diamonds} indicate
measurements using HARPS, ESPaDOnS, and PEPSI high-resolution
spectropolarimetric observations, respectively, and are based on a line 
mask with metal lines. The green dashed line is a sinusoidal fit to the 
$\bz{}$ values with a period
of 7.41\,yr.}
   \label{fig:Bevol}
\end{figure}

The new measurements of the mean longitudinal magnetic field values plotted 
together with older measurements discussed in the past by
\cite{Jarvinen2022}
are presented in Fig.~\ref{fig:Bevol} as a function of phase. As magnetic 
O-type stars are surrounded by magnetospheres, the observed dispersion of 
data points is usually rather large
\citep[e.g.][]{2015MNRAS.447.1885H,2015MNRAS.447.2551W}.

As already discussed in
\citet{Jarvinen2022}, 
the PEPSI spectra cover a much shorter wavelength region than HARPS spectra
and even shorter when compared to ESPaDOnS spectra. In comparison to 
previous PEPSI observations, the covered spectral region in the new PEPSI 
observation 
{  obtained in 2022} 
(see Section~\ref{sec:PEPSIobs} for more details) contains only eight useful 
metal lines. The low number of lines available in the PEPSI spectrum may 
explain the discrepancy between the latest ESPaDOnS and PEPSI 
longitudinal magnetic field measurements.

Our observations are in agreement with previous observations of massive O- 
and B-type stars showing that the magnetic field phase curves usually vary 
smoothly with a single wave related to a dipolar structure of the magnetic 
field geometry. Using all available magnetic field measurements covering 
now {  9.9}\,yr we reperformed a period analysis utilizing the 
Levenberg-Marquardt method 
\citep{Press1992}, 
which led to a stellar rotation period of $7.41\pm0.30$\,yr.

Only the 54\,yr rotation period of the magnetic star HD\,108
\citep{Rauw2023}
exceeds that of \hd{}. With the refined period, which within the error is 
the same as previously reported by
\citet{Jarvinen2022},
the data is fitted with a simple cosine curve illustrated with a dashed 
dark green line in Fig.~\ref{fig:Bevol}. From the variation of the 
longitudinal magnetic field over the rotation period, we determine an 
average magnetic field of $-290$\,G and a semiamplitude of 275\,G.

The effective magnetic field of an oblique dipole rotator is

\begin{equation}
B_e=\frac{1}{20} \frac{15+u}{3-u} B_p (\cos \beta \cos i +\sin \beta \sin i \cos \phi)
\label{stibbs}
\end{equation}

\citep{Stibbs1950,Preston1967}, 
where $B_p$ is the polar field strength, $u$ the limb darkening coefficient, 
$i$ the inclination, $\beta$ the obliquity, and $\phi$ the rotational phase. 
To determine the polar field strength from Equation~\ref{stibbs}, 
both the inclination and the obliquity angles would be needed. These angles 
can not be determined separately from the data. However, from 
Equation~\ref{stibbs} follows
\begin{equation}
 r = \frac{B_e^{\rm min}}{B_e^{\rm max}} = \frac{\cos(i+\beta)}{\cos(i-\beta)}.
\end{equation}
We can conclude that $i + \beta \gtrsim 90^\circ$ from the strong asymmetry 
of the magnetic field variation with respect to zero. From this follows that 
the dipolar field strength is larger than 1950\,G for $i= \beta = 
45^\circ$ and reaches values of more than 5\,kG for $i=10^\circ$ or 
$i=80^\circ$.

%
\section{H$\alpha$ and H$\beta$ variability caused by an extended magnetosphere}
\label{sec:Halpha}

In magnetic massive stars the observed \halp{} line emission is usually 
dominated by the magnetospheric wind material 
\citep[e.g.,][]{Petit2013,udd17}.
As discussed already in 
\citet{Hubrig2020},
the emission line profiles of \halp{} in \hd{} are highly variable. Both 
double- and triple-peak emission profiles were detected. In the observations 
with the high-resolution instruments used in this study, the \halp{} line is 
always included, with the exception of PEPSI. The new PEPSI observations 
include the \halp{} line, but do not cover the H$\beta$ line, while only 
the H$\beta$ line was obtained in observations carried out in 2020.

For massive stars with detected magnetic fields, magnetically trapped wind 
material frequently leads to rotationally modulated \halp{} emission.
In a previous work on \hd{},
\cite{Jarvinen2022} 
reported that the equivalent widths of the \halp{} and H$\beta$ lines show 
an increase during the most recent UVES and PEPSI observations, but further 
monitoring is necessary to confirm this trend. Further, according to
\cite{Hubrig2020}, 
the \halp{} lines in the spectra of \hd{} show variability on time-scales of 
weeks and even hours. It is possible that this variability is related to 
the wind or the immediate environment of the star. 
\cite{udd13} 
suggested that stochastic \halp{} scatter may originate from small 
differences in the amount of material trapped in the dynamical magnetosphere 
at any given time.
%
\begin{figure}
 \centering 
\includegraphics[width=\columnwidth]{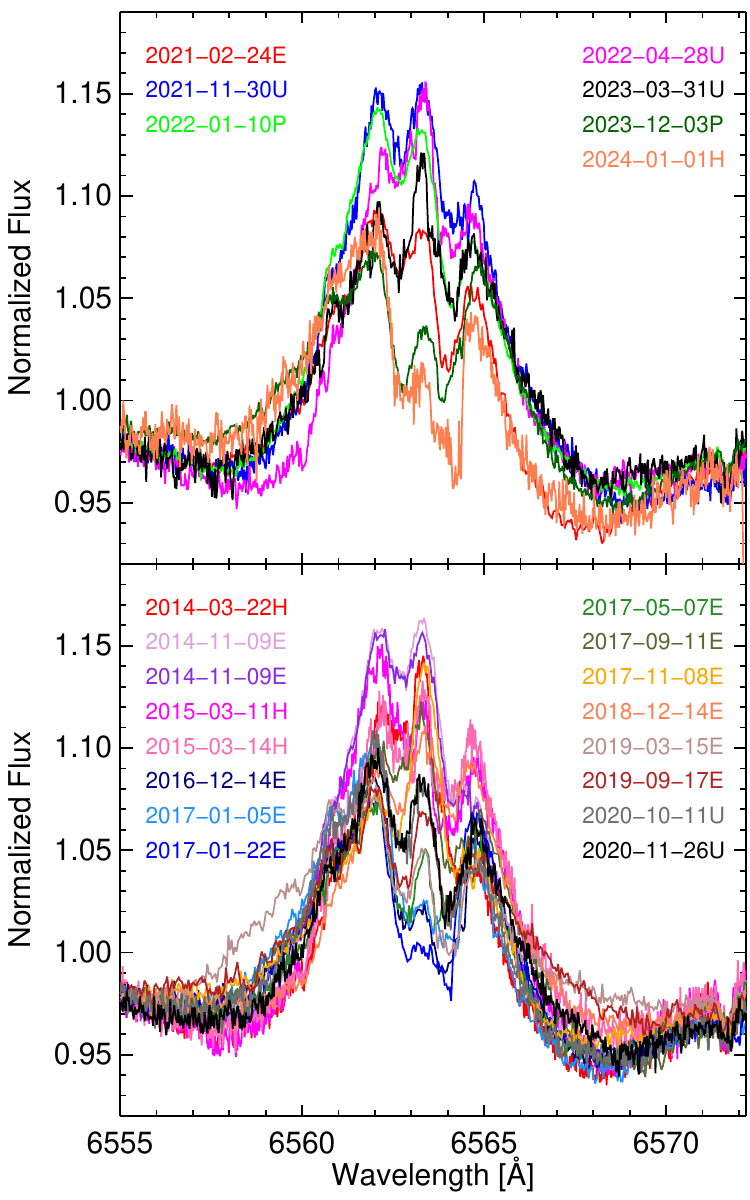}
\caption{
The overplotted \halp{} lines of \hd{}.
Each profile is plotted with a different color and the instrument used is
indicated after the data: HARPS (H), ESPaDOnS (E), UVES (U), and PEPSI (P).
{ 
The upper panel shows the profiles based on the new spectra and the bottom 
panel shows the profiles already published in
\cite{Jarvinen2022}.
}
         }
   \label{fig:Halphaprof}
\end{figure}
%
\begin{figure}
 \centering 
\includegraphics[width=\columnwidth]{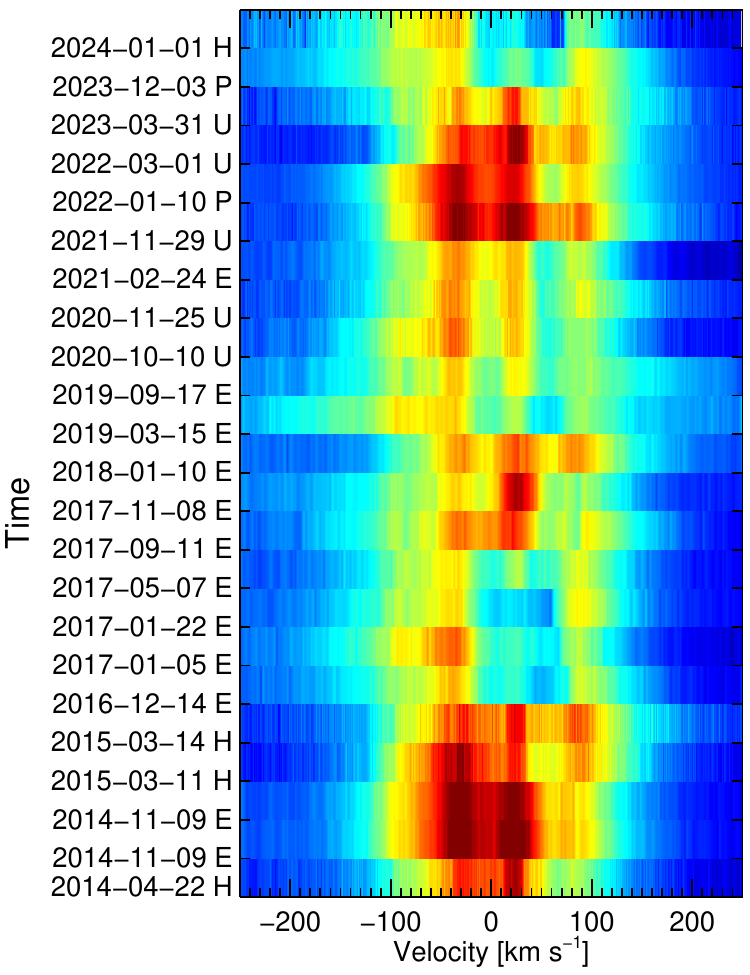}
\caption{
Dynamical spectrum showing the variability of the \halp{} profiles presented 
in Fig.~\ref{fig:Halphaprof}. The red colour corresponds to the strongest 
emission, while the blue colour shows the H$\alpha$ profile wings appearing 
in absorption. 
         }
   \label{fig:Halpha}
\end{figure}
%
\begin{figure}
 \centering 
\includegraphics[width=\columnwidth]{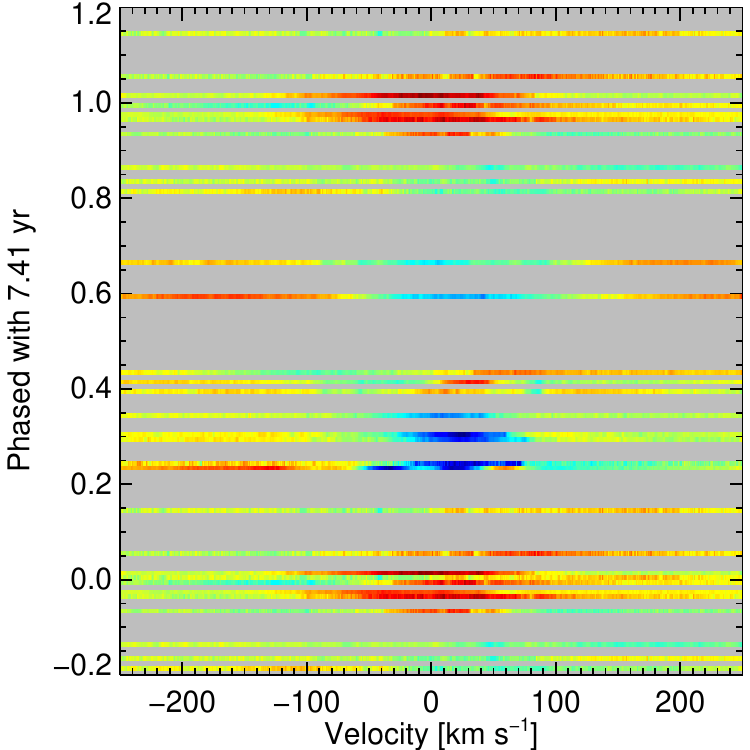}
\caption{
{ 
Dynamical spectrum of the differences of the individual \halp{} profiles with
respect to the mean profile phased with the period of 7.41\,yr.}
         }
   \label{fig:Halphadiff}
\end{figure}

In Fig.~\ref{fig:Halphaprof} we present all available high-resolution \halp{}
profiles overplotted and in Fig.~\ref{fig:Halpha} we display them as a 
dynamical spectrum with time. 
{ 
In Fig.~\ref{fig:Halphadiff} we present the differences of the individual 
\halp{} profiles with respect to the mean profile as a function of the phase. 
Both dynamical plots show, when combined with information about the magnetic 
field strength, that the \halp{} emission is the strongest at the best 
visibility of the negative magnetic pole.
%
\begin{figure}
 \centering 
\includegraphics[width=\columnwidth]{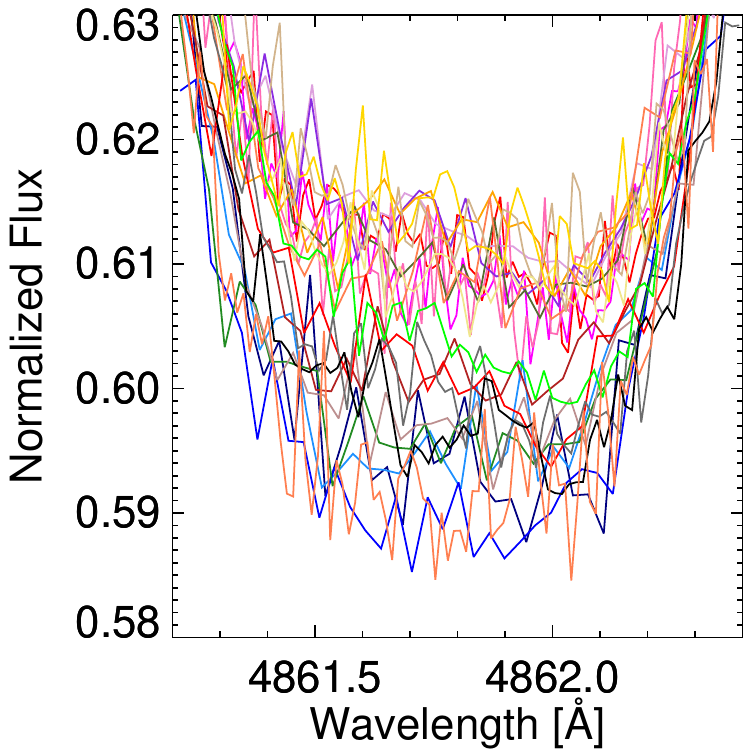}
\caption{
Zoom-in to the core of the H$\beta$ profiles. Each individual spectrum has 
been plotted with a different color.
         }
   \label{fig:Hbcoreprof}
\end{figure}

The behaviour of the hydrogen line H$\beta$ in \hd{} has not been studied in 
similar detail. Albeit the H$\beta$ profile does not show similar prominent 
changes as detected in \halp{}, we observe that both the width and depth 
of this line are variable. A zoom-in to the H$\beta$ line core is presented 
in Fig.~\ref{fig:Hbcoreprof}. In order to investigate if there is any 
correlation between the behaviour of H$\beta$ and \halp{} profiles, we 
present the H$\beta$ profiles as a dynamical spectrum in 
Fig.~\ref{fig:Hbeta}, which can be compared with Fig.~\ref{fig:Halpha}. The 
comparison shows that H$\beta$ line absorption cores are less deep in the 
phases when the \halp{} line profile exhibits stronger emission, indicating 
that the H$\beta$ lines are most probably filled-in with emission in these 
phases.
%
\begin{figure}
 \centering 
\includegraphics[width=\columnwidth]{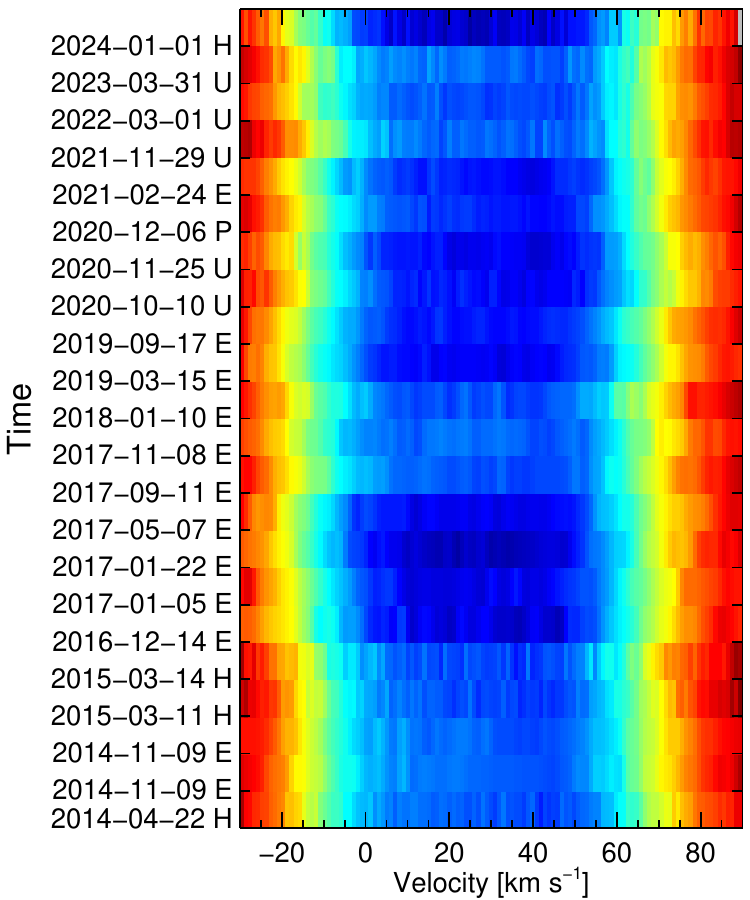}
\caption{
Dynamical presentation of the H$\beta$ core profiles shown in 
Fig.~\ref{fig:Hbcoreprof}.
The darkest blue color shows the deepest absorption whereas the redder the 
color the closer the profile is to the continuum.
         }
   \label{fig:Hbeta}
\end{figure}

The strong surface magnetic field and the relatively low mass loss rate of 
\hd{} lead to an extended magnetosphere, where the gas is trapped by the 
magnetic field. 
\cite{shenar2017} 
find a value of $10^{-9} M_\odot$\,yr$^{-1}$ for the mass loss rate if the 
magnetic field is not taken into account. The impact of the stellar magnetic 
field on the wind is described by the confinement parameter 
\citep{udd02},
\begin{equation}
  \eta_* = \frac{B_{eq}^2 R_*^2}{\dot{M}_{B=0} v\infty},
\end{equation}
where $B_{eq}$ is the magnetic field strength at the equator, $R_*$ the 
stellar radius; $\dot{M}_{B=0}$ is the mass loss rate  and $v_\infty$ the 
terminal speed found when the magnetic field is ignored. For \hd{}, using a 
dipolar field with a polar strength of 2\,kG and a terminal velocity of 
1700\,km\,s$^{-1}$, this yields a value of $1.7 \times 10^4$ and therefore 
a very extended magnetosphere. For a dipolar field, the Alfv\'en radius is  
given by the approximation 
\citep{udd08}
\begin{equation}
 \frac{R_A}{R_*} \approx 0.3 + (\eta_*+0.25)^{1/4}. 
\end{equation}
From this follows an Alfv\'en radius of 11.7\,$R_*$ for 
$\eta_*=1.7 \times 10^4$. The farthest closed field line crosses the 
equator at
\begin{equation}
  R_c \approx R_*+0.7(R_A-R_*).
\end{equation}
With $R_A = 11.7\,R_*$, we find $R_c=8.5 R_*$. 

The dynamical impact of the stellar rotation is given by the parameter 
\begin{equation}
 W = \frac{V_{\rm rot}}{V_{\rm orb}} = \frac{\Omega R_*}{\sqrt{GM_*/R_*}},
\end{equation} 
the ratio of the equatorial rotation speed and the Keplerian orbital speed 
at the stellar surface. With the rotation period of 7.41\,yr derived from 
the \halp{} variation, a value of $1.8 \times 10^{-4}$ follows, indicating a 
negligible impact of the stellar rotation. This is also manifested through 
the large Kepler corotation radius,
\begin{equation} 
 R_K = \left(\frac{G M_* }{\Omega^2}\right)^{1/3} = W^{-2/3} R_*,
\end{equation}  
for which we find a value of 313\,$R_*$. For the rotation to have a 
significant effect on the structure of the magnetosphere, the Kepler 
corotation radius must be smaller than the Alfv\'en radius. In this case, a 
centrifugal magnetosphere (CM) forms, where the gas corotates with the star. 
In contrast, in dynamical magnetospheres (DM) the Kepler corotation radius 
is larger than the Alfv\'en radius and the gas flows back to the star along 
the field lines of the magnetic field \citep{Petit2013}. In the case of \hd{} 
the Kepler corotation radius is much larger than the Alfv\'en radius and the 
magnetosphere is thus dynamical.

A comparison of the variations of the magnetic field and the \halp{} 
emission shows that they both follow the same cycle, i.e.\ the stellar 
rotation. From this we conclude that the star is an oblique rotator with a 
dipolar magnetic field that is tilted against the rotation axis. As the 
magnetic field is rather strong, a large fraction of the gas flowing from 
the star is trapped and forms a magnetosphere centered around the magnetic 
equator. This magnetosphere is the source of the \halp{} emission and the 
line width is determined by the velocity of the trapped gas, which is much 
lower than in the line driven wind along open field lines.

The radial velocity of \hd{} was first measured by 
\cite{Neubauer1943}, 
who found a value of $15.6\pm1.4$\,km\,s$^{-1}$. More recent measurements 
found values about twice as high. 
\cite{Boyajian2007} 
reports a value of $35.4\pm1.4$\,km\,s$^{-1}$ while 
\cite{Castro2015} 
found $29.5\pm1.0$\,km\,s$^{-1}$ and 
\cite{Hubrig2020} 
arrived at $27.0\pm0.1$\,km\,s$^{-1}$. Radial velocity of about 
30\,km\,s$^{-1}$ follows from dynamical presentation of the H$\beta$ core 
(Figure \ref{fig:Hbeta}), inline with the other recent determinations. If 
we correct for the radial velocity of the star, the side features in 
Fig.~\ref{fig:Halpha} show velocities of about 75\,km\,s$^{-1}$ and 
$-$75\,km\,s$^{-1}$. If these features were caused by an object in keplerian 
rotation around \hd{}, the orbital radius would have to be 78 stellar radii. 
An object in co-rotation with the star, on the other hand, would have to be 
located at 626 stellar radii to match a velocity of 75\,km\,s$^{-1}$. 
Neither assumption seems plausible.

A periodic \halp{} emission pattern has been modelled for a few other 
magnetic massive stars, $\theta^1$\,Ori\,C, HD\,191612, and HD\,57682:

$\theta^1$\,Ori\,C rotates with a period of 15.4\,d. The mass loss rate is 
$3.3 \times 10^{-7} M_\odot$\,yr$^{-1}$, and the polar field strength is 
1.1\,kG 
\citep{Stahl2008}.
The stellar magnetosphere was modelled in both 2D and 3D 
\citep{gagne05, udd13}.
The observed variation of the \halp{} line is reproduced in 
\cite{udd13} 
under the assumption that the magnetic field is tilted by 45$^\circ$ against 
the rotation axis and that the angle between the rotation axis and the 
viewing direction of the observer is also 45$^\circ$. The star and its 
magnetosphere are then seen pole-on (with respect to the magnetic field) at 
phase $\phi=0$ and edge-on at phase $\phi=0.5$. 

HD\,191612 is a magnetic O star with a rotation period of 538\,d, a mass 
loss rate of $1.6 \times 10^{-6}\,\rm{M}_\odot$\,yr$^{-1}$ and a polar 
magnetic field strength of $2450 \pm 400$\,G 
\citep{Howarth2007, Wade2011}.
\cite{Sundqvist2012} 
modelled the \halp{} emission of HD\,191612 based on numerical simulations 
of the stellar magnetosphere and a line-driven wind. The variation of the 
longitudinal magnetic field suggests a large obliquity $\beta=67^\circ$ and 
an inclination $i=30^\circ$ while the \halp{} emission was best reproduced 
with $\beta=i=50^\circ$. 

HD\,57682 is an O9 supergiant with a rotation period of 65.4\,d and a 
polar field strength of 880 G \citep{Grunhut2012}. With a mass loss rate 
of $1.4 \times 10^{-9}\,M_\odot$\,yr$^{-1}$, a terminal velocity of 
1200\,km\,s$^{-1}$ and a radius of 7\,$R_\odot$, the confinement parameter 
is $1.4 \times 10^4$. 
\cite{Grunhut2012} 
were able to model the observed \halp{} variation but had to assume a 
substantially larger value for the mass loss rate than the value inferred 
from UV by 
\cite{Grunhut2009}, 
contrary to the expectation of a mass loss rate lowered by the presence of a 
magnetic field 
\citep{udd08}. 
As the magnetic field variation is symmetric with respect to zero, a large 
obliqueness seems likely for this star.
%
\begin{figure}
\centering
\includegraphics[width=0.49\columnwidth]{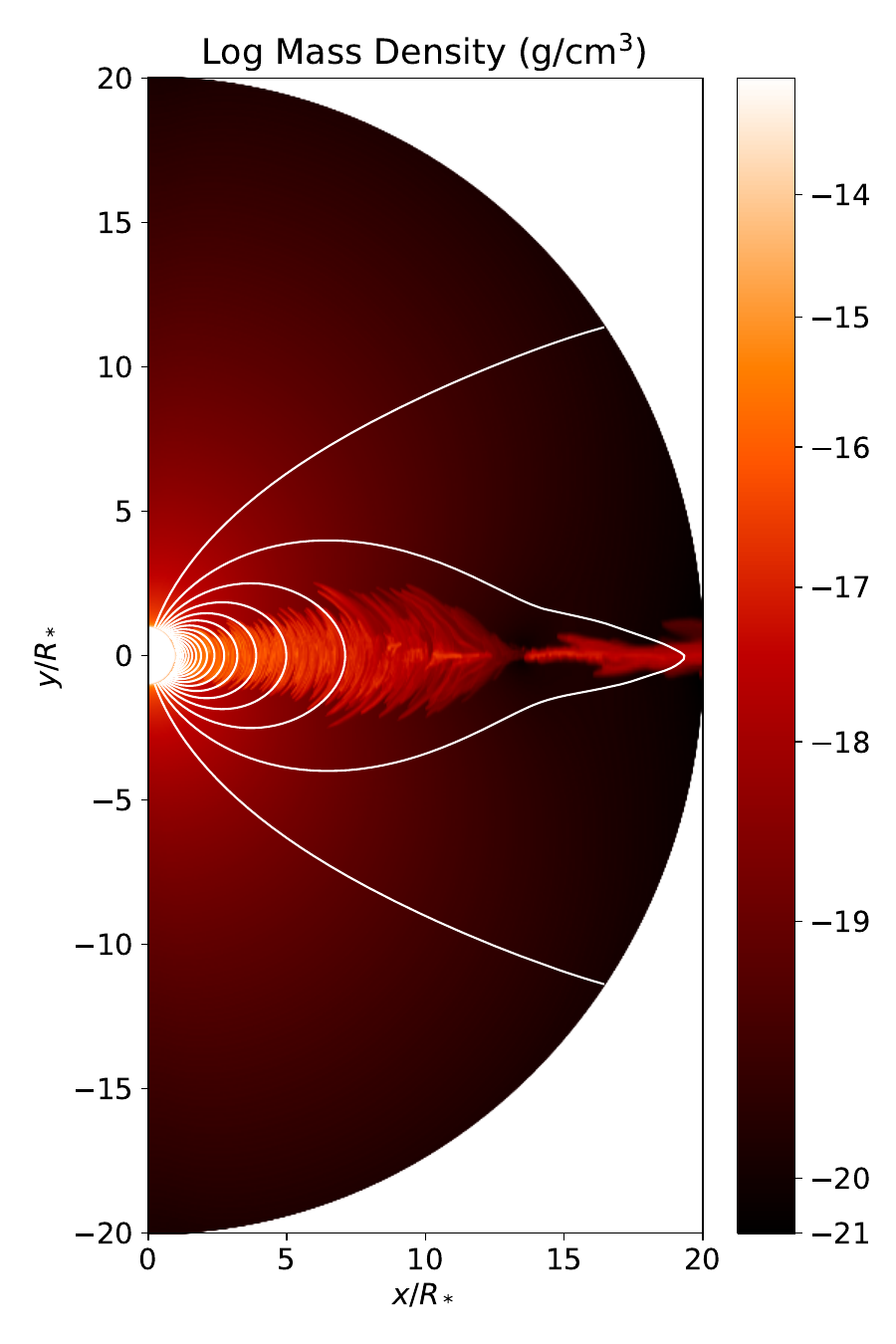}
\includegraphics[width=0.49\columnwidth]{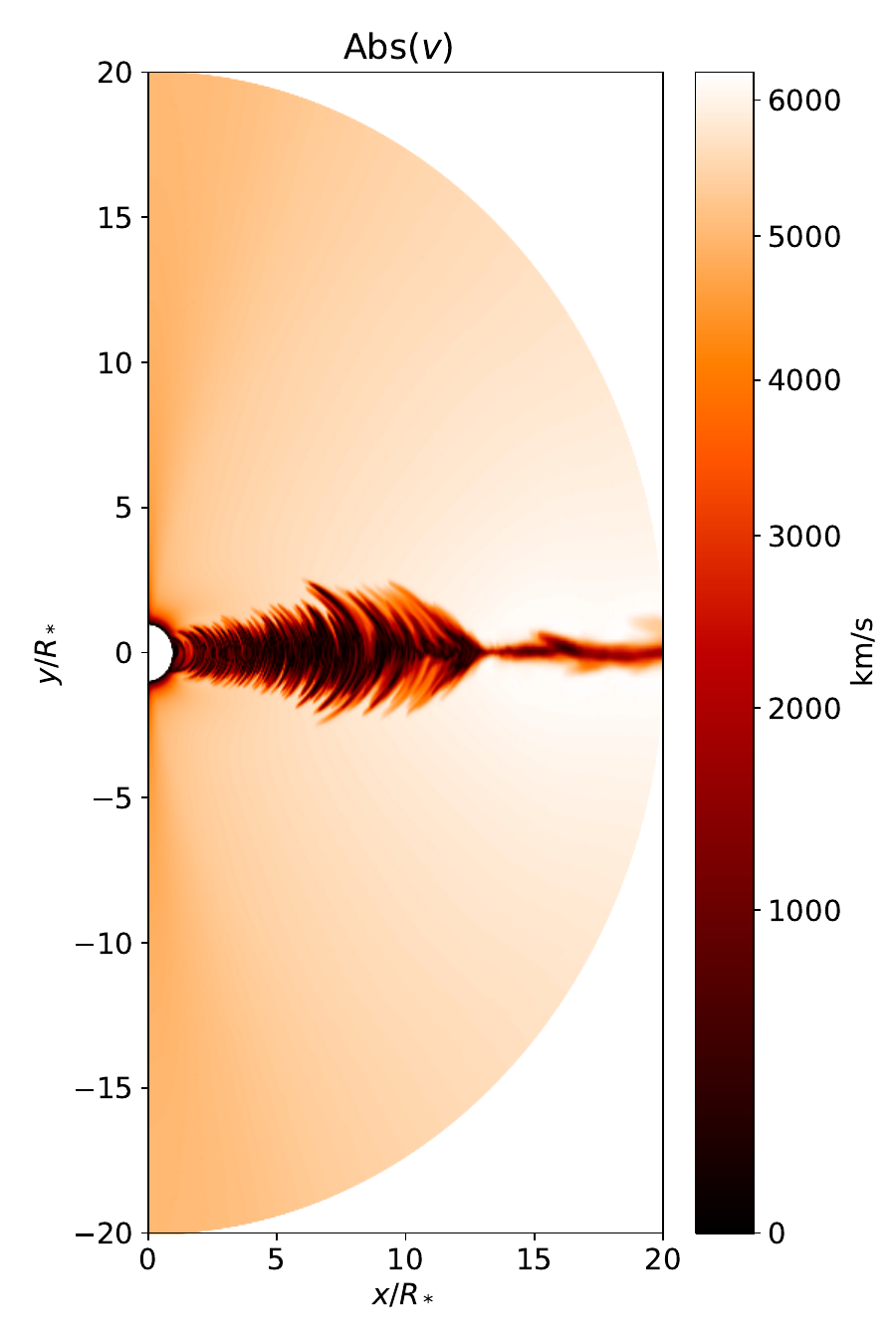}
\caption{Left: Mass density distribution and magnetic field lines from a 
numerical simulation of \hd{}. Right: Velocity amplitude.
}
  \label{fig:simulation}
\end{figure}
  
Figure \ref{fig:simulation} shows the distributions of mass density and gas 
velocity amplitude from a numerical simulation using the Nirvana MHD code 
\citep{ziegler04,ziegler05} 
with the setup described in 
\citet{Hubrig2019}. 
The Nirvana code solves the equation of motion, the induction equation, and 
the equations of energy and mass conservation for a fully compressible, 
ionised gas using a finite volume method. We have applied it for an 
axisymmetric configuration using spherical polar coordinates and an 
isothermal gas with a dipole field rooted in the star. Both plots display 
averages over 100 snapshots. The gas flowing from the star is mostly 
trapped by the magnetic field and accumulates in a disk-shaped region 
around the equatorial plane. The disk is separated into two regions. Within 
the Alfv\'en radius, it is rather thick and the gas moves back to the star 
in a zig-zag pattern where the motion is almost parallel to the magnetic 
field. This leads to a comb-like substructure of the inner disk. The outer 
part of the disk is much thinner and the gas moves outwards, dragging the 
magnetic field along. 
  
Our model has been tuned to meet the terminal velocity
and mass loss rate derived from observations assuming a weak magnetic field.
The presence of a strong magnetic field causes a decrease of the mass loss rate
and an increase of the terminal velocity.
The reduction of the mass loss rate can be explained by the trapping
of originally outflowing gas in the region of closed field lines.
The increased wind speed was first observed by \cite{udd02},
who attributed it to the divergence of the flow in the open field line region.
It is more pronounced in this work as \hd{} is an extreme case with its high confinement parameter.
The discrepancy between the velocity found in the simulations and the observed value
of the terminal velocity could be reduced by adjusting the wind force,
but that would require a large amount of CPU time and not change the structure
of the gas flow in the region of closed field lines.

As the simulation setup is isothermal, we can not model the \halp{} line 
and have to speculate about the origin of the emission. The Magnetically 
Controlled Wind Shock (MCWS) model by 
\cite{babel97a} 
predicts the formation of shock fronts in which the outflowing gas is 
heated up. The heated gas fills the volume between a disk in the equatorial 
plane and the shock fronts, which are located at mid-latitudes, and is the 
source of X-ray emission. 
\cite{babel97a} 
originally proposed the MCWS model for the A0p star IQ\,Aur with a mass 
loss rate between $10^{-11}$ to $10^{-10}$ $M_\odot$ / yr, a terminal 
speed $v_\infty=800$ km\,s$^{-1}$, a polar magnetic field strength 
$B_0$ = 4kG, and a radius of $5.1$ solar radii. From these parameters, a 
value between $10^6$ and $10^7$ follows for the confinement parameter, 
which justifies the assumption of an unperturbed dipole field. As 
Fig.~\ref{fig:simulation} shows, this is also fulfilled in our simulation 
up to a certain distance, after which the magnetic field is stretched away 
from the star and eventually opens. We can therefore expect the region where 
an increased density occurs to be the source of X-ray emission. The zig-zag 
structure is not predicted by the MCWS model, though.   

\cite{udd14} 
carried out numerical simulations of the wind confinement of a model star 
with parameters close to those of the O-type supergiant $\zeta$\,Pup and 
confinement parameters of 10 and 100. The simulations included heat 
transport and cooling and found a shock-heated region at low latitudes 
outside a radius of about two stellar radii. A zig-zag shaped gas flow back 
to the star is found in the region of closed field lines near the star, but 
it changes its direction less frequently than in our simulations. We 
attribute this difference to the higher value of the confinement parameter 
and the higher spatial resolution in our simulations.   

It should also be noted that as the magnetosphere of \hd{} is quite extended, 
its more distant part will never be completely obscured by the stellar disc, 
though the part closest to the star may be. Moreover, the gas in the 
equatorial disk outside the radius $R_c$ moves away from the star at speeds 
much lower than the wind speed above the disk. 
the left and right features in the dynamical \halp{} spectrum. In the high 
density region near the magnetic equator, the gas motion is predominantly 
perpendicular to the equatorial plane. This implies that only a weak Doppler 
shift would be observed when the disk is seen edge-on. 

The \halp{} variation of \hd{} is in agreement with the picture of an oblique 
dipole rotator, as derived from the magnetic field variation. The primary 
maximum at $\phi=0$ then corresponds to a pole-on view (with respect to the 
magnetic field and the magnetosphere) while the secondary maximum at 
$\phi=0.5$ coincides with the magnetic minimum, i.e.\ an edge-on view. We 
can thus rule out extreme values for the inclination angle as neither 
$i=0^\circ$, $\beta=90^\circ$ nor $i=90^\circ$, $\beta=0^\circ$ would 
produce the observed variation. On the other hand, the secondary maximum 
can not easily be explained as a mere projection effect, as the  
magnetosphere is presumably seen edge-on. A possible reason could be a 
significant deviation of either the magnetic field or the magnetosphere 
from symmetry with respect to the equator or from axisymmetry. We note 
that in the simulation shown, axisymmetry has been assumed but no symmetry 
with respect to the equatorial plane has been imposed and the gas 
distribution does deviate from symmetry on small scales. 

\section{Discussion} 
\label{sec:dis}

Both theoretical arguments and numerical simulations indicate the presence 
of an extended disk of increased gas density in the equatorial plane of the 
magnetic field, where gas from the line-driven stellar wind is trapped. 
This disk is usually assumed to be the origin of the observed \halp{} 
emission, which peaks together with the strongest line-of-sight magnetic 
field. Already a decade ago,  
\cite{udd13} 
reported that the magnetically trapped wind material flowing from a star 
and subsequent gravitational infall of this material yields a circumstellar 
density high enough to result in strong Balmer line emission. To interpret 
the \halp{} emission in the magnetic star $\theta^1$\,Ori\,C, the authors 
analyzed the rotational modulation of this line, using for the first time
fully 3D MHD simulations of the magnetic channeling and confinement of its 
radiatively driven stellar wind. The synthetic \halp{} profiles were 
computed by solving the formal integral of radiative transfer in a 
cylindrical coordinate system aligned toward the observer. Although the 
simulations reproduced quite well both the magnitude and phase of the 
observed rotational variation, the observed profiles appeared to exhibit 
an asymmetry about phase 0.5 that was not found in the applied numerical 
models.
\cite{udd13} 
concluded that such asymmetries may have their origin in, e.g., 
non-dipolar components of the magnetic field.

The rather clear splitting in the structure of the emission \halp{} profile 
observed at the phase of best visibility of the negative magnetic pole and 
the absence of a profile asymmetry similar to that detected in 
$\theta^1$\,Ori\,C make \hd{} an ideal target for the application of 3D MHD 
simulations to model its H$\alpha$ line emission. The presented distinct 
variability of this line with well separated emission peaks permits to 
follow the contribution of the gas disk accumulated along the magnetic 
equator and was not observed in any previously studied magnetic massive 
O-type star. Obviously, 3D modeling of the \halp{} line emission in \hd{} 
will significantly improve our understanding of magnetospheres and
channeled wind outflows from magnetic massive stars.


\section*{Acknowledgements}

{  We thank the anonymous referee for their comments.}

Based on observations made with ESO Telescopes at the La Silla Paranal 
Observatory under programme IDs~0102.D-0234(C), 0106.D-0250(A), 
0106.D-0250(B), 0108.D-0233(B), 0110.D-0103(A), 0110.D-0103(B), 
{  0112.D-2090(A), and 112.25MG.001}.

Based on data acquired with the Potsdam Echelle Polarimetric and 
Spectroscopic Instrument (PEPSI) using the Large Binocular Telescope (LBT) 
in Arizona. The LBT is an international collaboration among institutions in 
the United States, Italy, and Germany. LBT Corporation partners are the 
University of Arizona on behalf of the Arizona university system; Istituto 
Nazionale di Astrofisica, Italy; LBT Beteiligungsgesellschaft, Germany, 
representing the Max-Planck Society, the Leibniz-Institute for Astrophysics 
Potsdam (AIP), and Heidelberg University; the Ohio State University; and 
the Research Corporation, on behalf of the University of Notre Dame, 
University of Minnesota, and University of Virginia.
Based on observations collected at the Canada-France-Hawaii Telescope (CFHT),
which is operated by the National Research Council of Canada, the Institut
National des Sciences de l'Univers of the Centre National de la Recherche
Scientifique of France, and the University of Hawaii.


\section*{Data Availability}

The data obtained with ESO facilities will be available in the ESO Archive at
http://archive.eso.org/ and can be found with the instrument and
object name.
The ESPaDOnS data underlying this article are available in the CFHT Science 
Archive at https://www.cadc-ccda.hia-iha.nrc-cnrc.gc.ca/en/cfht/ and can be 
accessed with the instrument and object name.
The PEPSI data underlying this article will be shared on a reasonable
request to the corresponding author.
%
%
\bibliography{HD54879}
%
%
\end{document}